\documentclass[%
 aip,
 jap,%
 amsmath,amssymb,groupedaddress,
 reprint,%
]{revtex4-1}
\usepackage{graphicx}
\usepackage{dcolumn}
\usepackage{bm}

\draft 

\begin{document}


\title{Readout of superconducting nanowire single-photon detectors at high count rates} 



\author{Andrew J. Kerman, Danna Rosenberg, Richard J. Molnar, and Eric A. Dauler}
\affiliation{MIT Lincoln Laboratory, Lexington, MA 02420}


\date{\today}

\begin{abstract}
Superconducting nanowire single-photon detectors are set apart from other photon counting technologies above all else by their extremely high speed, with few-ten-ps timing resolution, and recovery times $\tau_R\lesssim$10 ns after a detection event. In this work, however, we identify in the conventional electrical readout scheme a nonlinear interaction between the detector and its readout which can make stable, high-efficiency operation impossible at count rates even an order-of-magnitude less than $\tau_R^{-1}$. We present detailed experimental confirmation of this, and a theoretical model which quantitatively explains our observations. Finally, we describe an improved readout which circumvents this problem, allowing these detectors to be operated stably at high count rates, with a detection efficiency penalty determined purely by their inductive reset time.
\end{abstract}


\maketitle 

\begin{figure}
\includegraphics[width=3.4in]{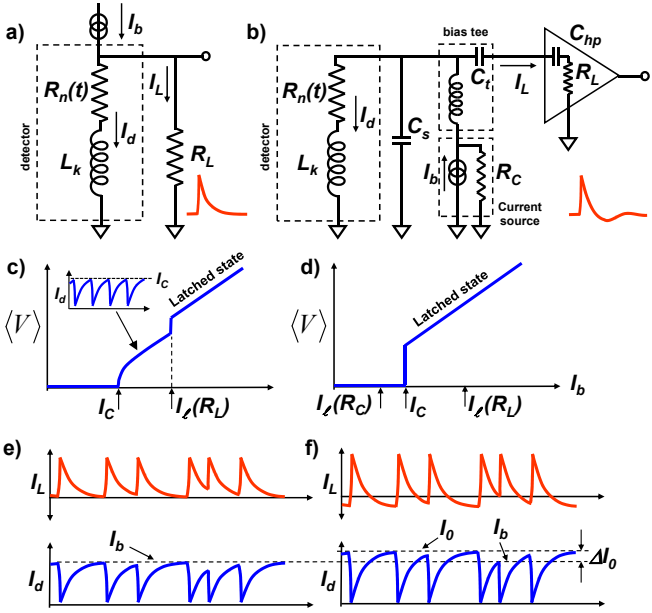}
\caption{\label{fig:circuits} Electrical circuits used to model SNSPDs and their readout. (a) Frequency-independent, resistive load, where the output pulse is a simple, asymmetric exponential. (b) typical experimental readout circuit with a bias tee, current source with finite compliance $R_C$, and AC-coupled amplifier. The high-pass corner frequency of the amplifier can produce an undershoot of the output pulse. The parasitic shunt capacitance across the detector due to its contact pads, wirebonds, etc..., shown in the figure as $C_s$, can nearly always be neglected since the time constant $R_LC_s$ is typically very short. (c) and (d) show the DC VI curves corresponding to the readouts of (a) and (b), assuming that the latching current $I_l(R_L)>I_C$. For the DC-coupled readout of (a), the wire undergoes persistent relaxation oscillations in the bias region $I_C<I_b<I_l$, and switches into the latched state when $I_b\ge I_l$. The AC-coupled circuit, however, cannot exhibit persistent relaxation oscillation, since at low frequencies the wire sees the DC compliance $R_C\gg R_L$ of the bias source, and the effective latching current is reduced from $I_l(R_L)$ to $I_l(R_C)<I_C$. Thus, when $I_C$ is transiently exceeded, only a short burst of relaxation oscillation occurs (which is not observed in a DC measurement), after which the device latches. (e) and (f) show the effect of high count rates. For the DC-coupled circuit of (a), the pulses always relax asymptotically towards a state with zero load voltage and $I_d=I_b$. For the typical experimental circuit of (b), however, the requirement that $\langle I_L\rangle=0$, $\langle I_d\rangle=I_b$ (due to the AC coupling) results in an increase $\Delta I_0$ in the detector current towards which the wire asymptotically recovers. Since $\Delta I_0$ is itself a function of the average pulse rate, the system becomes nonlinear.}
\end{figure}

Superconducting nanowire single photon detectors \cite{hadfield2012} (SNSPDs) show great promise to provide an unmatched combination of high efficiency and speed among shortwave-infrared photon counting technologies. They have separately demonstrated efficiencies at 1550nm of up to $\sim$90\% for single elements \cite{verma2012,*nam2012,pernice2012}, timing resolution down to a few ten ps \cite{dauler2008,najafi2012}, and count rates up to the GHz regime \cite{korneev2004,dauler2008,robinson2006,*robinson2007,*grein2011}. All of these attributes make SNSPDs attractive for many different applications, such as high-sensitivity optical communications \cite{robinson2006,*robinson2007,*grein2011}, quantum information processing \cite{acinDI2007,*aspuruphot2012,*obrien2012,*daulerQKD2010}, biomedical imaging \cite{biomed}, and quantum dot photonics \cite{stevens2008,*stevens2011,*waks2012,*correa2012}. However, in spite of the extensive experimental work to date \cite{hadfield2012}, only in a single, recent demonstration have all of these attributes been achieved simultaneously \cite{rosenberg2013}. The reason for this difficulty is that in nearly all cases there are engineering tradeoffs that require sacrificing performance in one attribute in order to optimize another. In this work, we focus on one such tradeoff associated with the detector readout, which has not previously been discussed. We identify and investigate a nonlinear interaction between the SNSPD and the circuit conventionally used to read it out, and show that in most cases the high count rates for which SNSPDs are well known cannot be achieved using this conventional readout scheme without substantial degradation of the efficiency and linearity. We then describe a modified cyrogenic readout that allows the efficiency and count rate to be decoupled, and linearity to be maintained to much higher count rates.

An SNSPD is often modeled using the simple circuit of Fig. \ref{fig:circuits}(a): the nanowire is represented by a kinetic inductance $L_k$ in series with a time varying hotspot resistance $R_n(t)$ (which is zero in the static, superconducting state). The usual readout involves coupling the wire directly to a transmission line, shown as a load resistance $R_L$. The device is biased with a static current $I_b$, which is close to but less than the nanowire's critical current $I_C$ (typically $\sim$10$\mu$A). When the wire absorbs a photon, a non-superconducting, resistive domain (also known as a hotspot) is formed, with a probability that can be near unity for $I_b\rightarrow I_C$ (this probability in most cases decreases exponentially for $I_b$ far below $I_C$). The hotspot quickly grows in size and total resistance due to Joule heating, forming a current divider with the load $R_L$. Reduction in the device current $I_d$ by this divider is opposed, however, by an induced voltage in the kinetic inductance, for a characteristic time $\sim L_k/R_n(t)$, during which the Joule heating causes $R_n(t)$ to increase to a value much larger than $R_L\approx 50\Omega$. Because of this large $R_n$, the current eventually diverts almost completely into the load, shutting off the Joule heating and allowing the detector to cool back down and return to the superconducting state. The current then returns to the detector over a characteristic time $\tau_R=L_k/R_L$, known as the reset time, resulting in the asymmetric pulse shape shown in Fig. \ref{fig:circuits}(a) \cite{kermanKI2006}. This self-resetting behavior of SNSPDs can be viewed as (unstable) negative electrothermal feedback to the wire by the shunt resistance $R_L$ \cite{kermanETF2009,yang2007}.

Quasi-static VI curves for the configuration of fig.~\ref{fig:circuits}(a) are shown schematically in fig.~\ref{fig:circuits}(c). They do not exhibit the usual, simple behavior of a superconductor in which it switches suddenly into the resistive state at $I_b=I_C$. Instead, due to the electrothermal feedback the VI curve exhibits two distinct non-superconducting regions. First, for $I_C<I_b<I_l$, where $I_l$ (to be described below) is known as the latching current, the wire's average resistance increases in a continuous, nonlinear fashion. This corresponds to a steady state of relaxation oscillation, in which the detector emits a periodic train of output pulses like those described above \cite{hadfield2005,kermanETF2009} [fig.\ref{fig:circuits}(c)], in which a pulse occurs each time $I_d$ crosses $I_C$ (in recovering towards $I_b>I_C$). As $I_b$ is turned up further above $I_C$, the frequency of these pulses correspondingly increases, producing the continuous increase in the average voltage across the load in the VI curve. In addition, it turns out that the (unstable) feedback associated with the oscillations also becomes more stable as $I_b$ is increased, until eventually at $I_b=I_l$ the wire ``latches" into a stable, resistive state known as a self-heating hotspot \cite{kermanETF2009}. In this latched state, a finite length of wire is maintained stably at a temperature above the critical temperature $T_C$ by a balance between Joule heating and cooling by conduction, stabilized by electrothermal feedback. The resistance of this spot is given approximately by: $R_{hs}\approx R_L(I_b/I_{SS}-1)$, where $I_{SS}$ is a constant determined by the heat transfer out of the wire \cite{kermanETF2009}.

While fig.~\ref{fig:circuits}(a) and the associated discussion is useful to qualitatively understand SNSPD behavior, the readout and biasing used in typical experiments is more correctly described by fig.~\ref{fig:circuits}(b): A bias tee is used to inject $I_b$ into the device, while directing its high-speed output to an AC-coupled low noise amplifier. The current bias is shown with a parallel resistance $R_C$, to account for its finite compliance. The input capacitance of the amplifier $C_{hp}$ will typically be much smaller than that of the bias tee, so it will dominate the series capacitance seen by the device (up to frequencies whose wavelengths are comparable to or longer than the electrical length between device and amplifier). This AC coupling turns out to have several important effects, which we now enumerate.


\begin{figure*}
\includegraphics[width=6.5in]{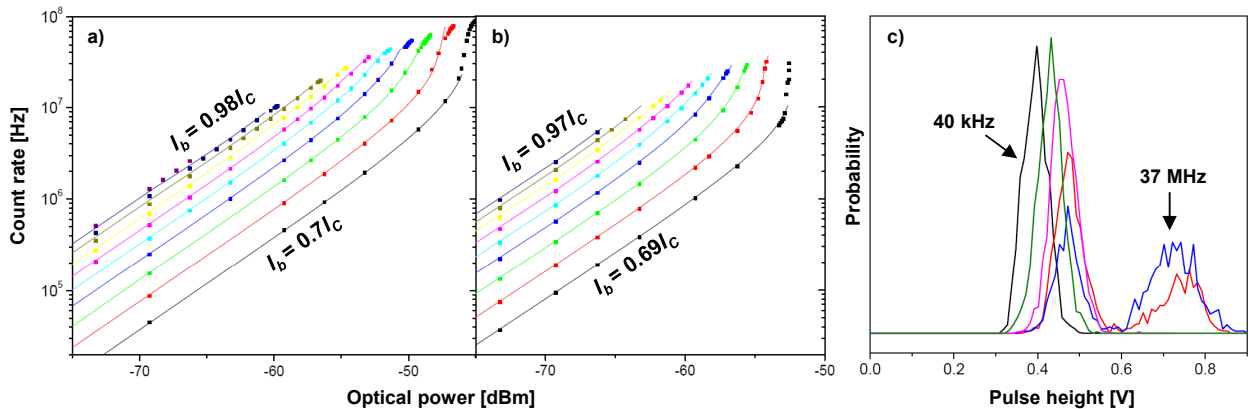}
\caption{\label{fig:ACbias} Output of AC-coupled circuit of fig.~\ref{fig:circuits}(b) vs. input photon flux. Panels (a) and (b) are for wires with different $L_k$, corresponding to $\tau_R=3.4$ nH and $\tau_R=7.2$nH, respectively; the different curves in each panel are for different bias currents, from bottom to top: $I_b/I_C=0.7,0.73,0.76,0.79,0.82,0.85,0.89,0.92,0.95,0.98$ (a) and $I_b/I_C=0.69,0.72,0.75,0.78,0.81,0.84,0.88,0.91,0.94,0.97$ (b). At low count rates, all of the curves are linear; however, starting at count rates as low as $\sim(20\times\tau_R)^{-1}$, the curves start to bend upward due to the nonlinearity discussed in the text. Solid lines are predictions of our model, based on independent measurements, with no fitting parameters. Panel (c) shows histograms of the measured electrical pulse \textit{height} for the same device as panel (b), at count rates of $0.04,3.1,6.7,24,37$ MHz. As described in the text, the average pulse height actually increases with count rate (the peak in the histogram shifts to the right) due to the nonlinearity. The separated peak at much higher amplitude that appears for the two highest count rates is associated with the onset of relaxation oscillation; discrepancies in (a) and (b) at the very highest count rates between the data and our predictions are likely a result of this onset, which is not included in our model.}
\end{figure*}

The first of these effects is evident in the quasi-static VI curve, shown schematically in fig.~\ref{fig:circuits}(d): there is no middle branch to the curve, and the detector appears to switch suddenly into the resistive state at $I_b=I_C$. In fact, when $I_C$ is first exceeded, the nanowire begins transiently to exhibit relaxation oscillations as above; however, the resulting nonzero average voltage across the wire begins to also appear across $R_C$ after the appropriate time constant associated with the bias tee (equivalently, at low enough frequencies the impedance looking out from the detector is $R_C$). As described above, the value of $R_n$ to which the negative feedback will tend to stabilize the wire increases with the load resistance; since the primary limitation on the speed of the electrothermal feedback is the inductive time constant of the circuit, larger $R_n$ then means faster feedback, a more stable hostpot, and a lower $I_l$. So, once the device starts to ``see" a large $R_C$, the effective latching current $I_l(R_C)$ becomes less than $I_b$, and the wire latches. The result is that on an oscilloscope one can observe a short burst of chirped oscillations (for a characteristic time mostly determined by the bias tee, typically $\sim\mu$s) followed by a \textit{static}, latched solution where: $R_n=R_C(I_b/I_{SS}-1)$, $I_d=I_{SS}$, and a current flowing through $R_C$ of $I_b-I_{SS}$.

Another consequence of the AC coupling shown in fig. \ref{fig:circuits}(b) is that the blocking capacitor forms part of a damped, series $LRC$ oscillator, with quality factor $Q=\sqrt{\tau_R /R_LC_{hp}}$. Photon detection events act as impulse perturbations to this resonant circuit. If $Q<1$ it is overdamped and the trailing pulse edge will be monotonic as shown in fig. \ref{fig:circuits}(a); if $Q>1$, however, the circuit will exhibit ringing in response to each detection, and the current through the device will overshoot $I_b$ on the trailing edge of each pulse, as shown in fig. \ref{fig:circuits}(b). If $I_b$ is close to $I_C$, the device can transiently exceed $I_C$ after the very first pulse and switch into the latched state as discussed above. This produces an apparently lower $I_C$ than the true value, and makes the highest detection efficiencies inaccessible. In addition, for currents near $I_C$ this can produce apparent afterpulsing due to the pulse overshoot. Attenuation between the device and the amplifier can mitigate these effects, but only at the expense of a reduced signal-to-noise ratio (which can increase the timing jitter).

The final consequence of the AC coupling occurs when the detector is firing at a high average rate, as illustrated in figs. \ref{fig:circuits}(e) and (f). In this regime, at the output of the amplifier the AC coupling results in an offset from zero of the apparent ``background"  load current (signal voltage) to which the signal returns between pulses. This seems innocuous enough; however, if we note that necessarily $\langle I_L\rangle=0$, $\langle I_d\rangle=I_b$ (looking out from the current source, there is no other DC path to ground but through the nanowire), we arrive at the conclusion that the current pulses in the nanowire at high count rates must recover towards a background value $I_0$ \textit{greater than} $I_b$, such that the detection efficiency towards which the detector recovers is $P_D(I_0)>P_D(I_b)$. Since the effective offset current $\Delta I_0\equiv I_0-I_b$ (and the corresponding change in detection efficiency) is itself a function of the average detector count rate $R_{obs}$, we now have a nonlinear system \cite{flatness}. As the detector count rate is increased towards $\tau_R^{-1}$, the nonlinearity can become very strong.

Figure~\ref{fig:ACbias} shows our observations of this effect, for two detectors (having different $L_k$) as a function of incident continuous-wave optical power, and for several different $I_b$ approaching $I_C$. At low count rates where $R_{obs}\ll\tau_R^{-1}$, all of the curves are linear with unit slope on the log-log plot (a characteristic of any single-photon detection process), and each has a vertical position given by the corresponding detection efficiency. However, starting at count rates as low as $\sim(20\times\tau_R)^{-1}$, the curves become nonlinear and bend upward (note that this effect was observed in ref.~\onlinecite{robinson2006,*robinson2007,*grein2011}, but not understood at the time). Due to the nonlinearity in this regime, the detector will have an effective average detection efficiency, timing jitter, and dark count rate \textit{which all depend on the instantaneous input photon rate} (averaged over the time constant of the bias tee); this can be highly problematic, particularly in communications applications, where the pulse pattern contains information. Note also that the nonlinear effect sets in at relatively low count rates: for the devices in figs.~\ref{fig:ACbias}(a) and (b) with $\tau_R=$3.4 ns and 7.2 ns, it starts to become important already at count rates of 20 MHz and 10 MHz, respectively. These rates are more than an order of magnitude smaller than $\tau_R^{-1}$, showing that using the conventional readout of fig.~\ref{fig:circuits}(b), in general one cannot infer high count rate ($\sim\tau_R^{-1}$) capability for SNSPDs purely from low-flux reset time measurements.

The nonlinear effect can also be seen from a different point of view: the \textit{amplitude} of the output pulses. As $R_{obs}$ and $I_0$ increase, the signal swing of each pulse increases as well (to $\sim I_0R_L>I_bR_L$), as illustrated in fig.~\ref{fig:circuits}(f). Figure \ref{fig:ACbias}(c) shows histograms of the observed pulse heights for a single detector as a function of observed count rate (the incident power is varied), from $R_{obs}=$40 kcounts/s to $\sim$40 Mcounts/s. As the count rate is increased, the center of the distribution indeed shifts to larger values. Eventually, a second peak appears at larger amplitude, which corresponds to the same relaxation oscillations discussed above in the context of $I_b>I_C$; however, in this case it is $I_0$ that has increased above $I_C$ due to the nonlinearity. As the input power is yet further increased, the detector eventually latches, just as in the case of $I_b>I_l$; this is reflected in the termination of all sets of data at some maximum input power and output count rate. Note that this happens sooner for the curves corresponding to larger $I_b$ near $I_C$, such that the highest average count rates can only be achieved by turning $I_b$ down significantly ($\sim0.7I_C$) and letting the nonlinearity increase the effective bias $I_0$ up to near $I_C$.

\begin{figure}
\includegraphics[width=3.25in]{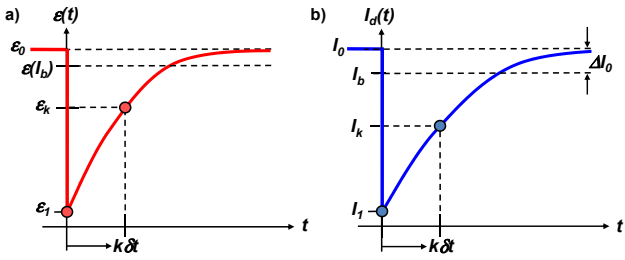}
\caption{\label{fig:reset} Illustration of the quantities used in calculating the self-consistent weighted average over a Poisson photon arrival time distribution.}
\end{figure}

We now construct a quantitative model for this nonlinearity, building on the method of ref.~\onlinecite{rosenberg2013}, by evaluating the self-consistent time average of the detection rate $R_{obs}$ over a Poissonian photon arrival time distribution at arbitrary input photon rates $R_{in}$. To do this, we discretize the device's output pulse shape following a detection at $t=0$ into bins of duration $\delta t\ll\tau_R$, starting from an instantaneous current $I_0$ as illustrated in fig. \ref{fig:reset}, with corresponding detection efficiency $\epsilon_0$ and dark count rate $d_0$. For the $n^\textrm{th}$ time bin at $t=n\delta t$, we define an associated current $I_n$ and the corresponding detection efficiency $\epsilon_n\equiv P_D(I_n)$ and dark count rate $d_n\equiv R_{dark}(I_n)$, where $P_D(I)$ and $R_{dark}(I)$ are the measured detection efficiency (in the $R_{in}\rightarrow0$ limit) and dark count rate as a function of bias current \cite{dark}. The output count rate can then be written \cite{rosenberg2013}:

\begin{equation}
R_{obs}=R_{in}\epsilon_0+\;d_0-\sum_{n=1}p_n\left [R_{in}(\epsilon_0-\epsilon_n)+(d_0-d_n)\right]\label{eq:robs}
\end{equation}

\noindent where $p_n$ is the average probability that the detector samples the $n^\textrm{th}$ time interval of its reset, or equivalently: the conditional probability (averaged over many detection events) that the \textit{next} detection event does not occur until \textit{after} a time $n\delta t$ has elapsed. We define it recursively as:

\begin{equation}
p_n=p_{n-1}\left[1-(R_{in}\epsilon_n+d_n)\delta t\right]\label{eq:pn}
\end{equation}

\noindent with $p_1\equiv R_{obs}\delta t$. Note that $\epsilon_n$ and $d_n$ both depend implicity on $R_{obs}$, since $R_{obs}$ determines $I_0=I_b+\Delta I_0$ via the constraint: $\langle I_d\rangle=I_b$. Higher detection rates $R_{obs}$ cause $I_0$ to increase, changing $\epsilon_n$ and $d_n$. We solve equation \ref{eq:robs} numerically, using as input the measured $L_k$, $P_D(I)$ and $R_{dark}(I)$, and the results are shown by solid lines in fig. \ref{fig:ACbias}. The two panels are for two wires with different $L_k$, and the excellent agreement with our measurements is obtained without any adjustable parameters.

\begin{figure*}
\includegraphics[width=7in]{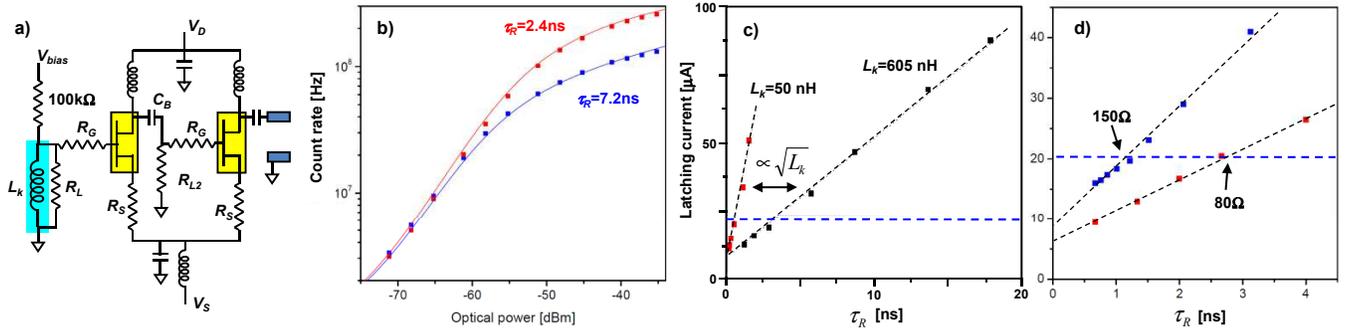}
\caption{\label{fig:DCcoupled} DC-coupled SNSPD readout. (a) schematic: the first stage drives the load $R_{L2}$, and $C_B$ in between the stages allows for simplified biasing; the circuit has a high-pass corner frequency of $\sim20$MHz. The resistors $R_G\sim 5\Omega$ and $R_S\sim 10\Omega$ are used to improve high-frequency stability. Typical bias parameters at $T=2.5K$ are $I_{DS}\sim2$mA, $V_D=0.7$V, $V_S=0.35$V ($\sim$0.7 mW power dissipation), which gives a typical gain of $\sim5-10$dB. (b) output count rate of the circuit when used with a device having $L_k=360$nH, and either $R_L=50\Omega$ (blue) or $R_L=150\Omega$ (red) (corresponding to $\tau_R=7.2$ns and $2.4$ns, respectively). The solid lines are from our model. Panels (c) and (d) illustrate the limitations on $R_L$ imposed by latching, as a function of $L_k$ and for two different substrate materials, respectively. (c) latching current $I_l(R_L)$ vs. $\tau_R$ (varied by changing $R_L$); the observed dependence is close to linear, as predicted in ref.~\onlinecite{kermanETF2009}. The horizontal dashed line is the critical current, and the maximum speed (without reduced efficiency) is where $I_l$ crosses $I_C$. Also consistent with ref.~\onlinecite{kermanETF2009} is the $\propto\sqrt{L_k}$ dependence shown in (c) of the slope of the line, indicating that if $R_L$ can be adjusted freely, the maximum speed only decreases as $\propto\sqrt{L_k}$ instead of $\propto L_k$ as would be the case for fixed $R_L$. Finally, in (d) we show similar data for two devices made using NbN grown on different substrates. The upper (blue) data is for NbN on sapphire, and the lower (red) data is for NbN on oxidized Si. The latching currents are significantly higher for the device made on sapphire (and correspondingly it can be made $\sim2\times$ faster than the one on SiO$_2$ without latching), indicating that the heat transfer between wire and substrate is significantly better in that case.}
\end{figure*}

So far we have illustrated several problems that can result from the conventional, AC-coupled readout circuit used with SNSPDs [c.f., fig.~\ref{fig:circuits}(b)]. Figure~\ref{fig:DCcoupled}(a) shows our simple solution to these problems: a preamplifier whose input stage is DC-coupled, and which presents to the nanowire a nearly frequency-independent, resistive impedance to ground. We have implemented this circuit using GaAs high-electron-mobility transistors (HEMTs) (Fujitsu FHX45X), where the load impedance at the gate is provided by a chip resistor intended for use at microwave frequencies (in our case, Mini Systems MSTF 2AN series). The circuit is a simple, two-stage, common source amplifier; since the detector is biased at ground, the source must be brought to positive voltage to turn the channel on. To decouple the supply voltages, both capacitors and inductors are used. To support the wide bandwidths required here, we use conical inductors (Coilcraft BCL-122JL), which have extremely high self-resonant frequencies, well out of our signal band. All of the circuit components are located on an Aluminum Nitride substrate, placed near the detector chip and indium soldered to a copper heat sink; wirebonds are used to connect the circuit to the detector and to its two supply voltages $V_D$ and $V_S$). Figure ~\ref{fig:DCcoupled}(b) shows our observations of count rate vs. incident optical power using this circuit, as well as the predictions of our model for the DC-coupled circuit (where we set $I_0=I_b$ as in ref.~\onlinecite{rosenberg2013}), which are in very good agreement. The detector now performs exactly as expected based on its reset time measured at low flux \cite{jitter}. 

Since the impedance $R_L$ seen by the device is determined by a chip resistor in this circuit, it can be increased without the constraint of any line impedance. Larger $R_L$ is desirable because it provides both a larger output signal and a faster reset time. However, as $R_L$ is increased, the latching current $I_l(R_L)$ decreases, eventually becoming less then $I_C$. Beyond this point the device can no longer be biased near the critical current and the maximum detection efficiency cannot be reached even at low count rates \cite{kermanETF2009}. Figure~\ref{fig:DCcoupled}(c) shows this effect, measured using our DC-coupled circuit, for detectors with two difference inductances. The reset times are varied for both of these detectors by changing $R_L$ at the input of the circuit. As discussed in ref.~\onlinecite{kermanETF2009}, the latching current decreases approximately linearly with $\tau_R$, with a slope $\propto\sqrt{L_k}$. This means that when $R_L$ can be freely adjusted to its maximum value where $I_l(R_L)=I_C$, the minimum reset time of the detector only scales as $\propto\sqrt{L_k}$, instead of linearly with $L_k$ as in the case of a fixed $R_L$.

Figure~\ref{fig:DCcoupled}(d) shows similar measurements of $I_l(R_L)$ for two detectors with the same $L_k\approx 200$nH, one made from NbN grown on sapphire and the other NbN grown on an oxidized silicon substrate, where the latter detector exhibits a significantly lower latching current $I_l$ for a given reset time $\tau_R$. Since NbN grown on sapphire is expected to produce better thermal contact (to phonons in the substrate) than for NbN grown on amorphous SiO$_2$, this result is in qualitative agreement with the expectation of ref.~\onlinecite{kermanETF2009}: that poorer heat transfer tends to stabilize the hotspot and lower $I_l$ \cite{heattrans}. On the other hand, cooling to the substrate that is \textit{too efficient} would be expected to reduce the detection efficiency, if the energy deposited by a photon is removed so quickly that a hotspot is never fully formed in the first place. This may constitute a new engineering tradeoff in SNSPDs, between high speed (in the form of weak latching) and high efficiency (in the form of a weak thermal link with the substrate). Future advances in materials and substrates for SNSPDs may soon have to address this if significant performance improvements are to be obtained.

A final advantage of the DC-coupled circuit used here is that it makes it easy experimentally to determine whether a detector is latching or not (and correspondingly if it can in fact be biased all the way up to its full critical current). Since it is effectively a realization of the simple circuit of fig.~\ref{fig:circuits}(a), the associated discussion applies: if $I_l>I_C$, one will see a region of bias current where the detector undergoes persistent relaxation oscillations, easily observable directly on an oscilloscope, or as a middle branch to the VI curve as in fig.~\ref{fig:circuits}(c). If $I_l<I_C$, however, this region will not be observed. By contrast, with the usual AC-coupled circuit of fig.~\ref{fig:circuits}, a sudden jump to a resistive state is observed in the VI curve \textit{in either case}, and it can be difficult to observe the short, transient burst of pulses which indicates $I_l>I_C$.

In conclusion, we have shown that in most cases the full potential for high-speed operation often associated with SNSPDs cannot be realized using the conventional readout circuit, due to a nonlinear interaction between this circuit and the detector. Our model for this interaction, when combined with independent device measurements, quantitatively explains our observations. We have demonstrated a simple cryogenic preamplifier circuit that decouples the detector from its circuit environment, and allows it to perform at count rates and efficiencies up to the theoretical limits imposed by its inductive reset time. This circuit is compact and relatively low power, and can readily be implemented for tens of detector channels using standard microwave hybrid circuit techniques. Scaling to hundreds or thousands of channels will likely require a more integrated solution, of smaller size and with lower power dissipation.

This work is sponsored by the Assistant Secretary of Defense for Research \& Engineering under Air Force Contract \#FA8721-05-C-0002.  Opinions, interpretations, conclusions and recommendations are those of the author and are not necessarily endorsed by the United States Government.

\bibliography{ACbias5}

\begin{thebibliography}{29}
\bibitem [{dar()}]{dark}%
  \BibitemOpen
  \href@noop {} {}\bibinfo {note} {Note that $R_{dark}(I)$ only then makes
  sense for $I_b$ far enough from $I_C$ that the total dark count rate is still
  much less than $\tau_R^{-1}$.}\BibitemShut {Stop}%
\bibitem [{jit()}]{jitter}%
  \BibitemOpen
  \href@noop {} {}\bibinfo {note} {One caveat to this is that since the timing
  jitter increases quickly as the bias current is reduced below $I_C$
  \cite{najafi2012}, the average jitter will be increased if
  $R_{obs}\sim\tau_R^{-1}$ and the detector current does not have time to
  recover fully after each detection event.}\BibitemShut {Stop}%
\bibitem [{hea()}]{heattrans}%
  \BibitemOpen
  \href@noop {} {}\bibinfo {note} {Note that heat transfer may also be affected
  by film \textit{thickness}, since phonon propagation time can also be
  important.}\BibitemShut {Stop}%
\end{thebibliography}

\end{document}